Growth of germanium-silver surface alloys followed by in-situ scanning tunneling microscopy: absence of germanene formation.

K. Zhang,[1] R. Bernard,[1] Y. Borensztein,[1] H. Cruguel,[1] G. Prévot[1]

[1]*Sorbonne Université, CNRS, Institut des NanoSciences de Paris, INSP, F-75005, Paris, France*

Abstract

Theoretical studies have shown that new physical properties such as tunable gap opening or quantum spin Hall effect could be expected from group IV graphene analogues (silicene, germanene, stanene). While there have been numerous studies of growth of such Si, Ge, Sn monolayers, the demonstration of their hexagonal organization has been often based on post-growth characterization and their analogy to graphene has remained controversial. Our real-time scanning tunneling microscopy (STM) observation during Ge deposition on Ag(111) in the [380 K - 430 K] temperature range reveals that Ag atoms are involved in all the structures observed before the formation of a second layer, rejecting the possible formation of germanene on this substrate within these experimental conditions. The observation by STM of Ge atomic diffusion shows that easy exchange between Ag and Ge atoms is responsible for the Ge-Ag surface alloying at such temperatures.

1. Introduction

Germanene, the two-dimensional allotrope of germanium, has been widely studied in the last decade, since initial works based on density functional theory (DFT) [1,2] have predicted that Ge could exist in a free-standing (FS) metastable configuration, where Ge atoms are hexagonally organized. Such germanene would possess an electronic structure showing Dirac cones at the Fermi level, and a high electrical conductivity. Experimentally, due to the lack of a graphitic form of germanium, which would permit its synthesis by exfoliation, germanene needs to be synthesized by deposition on a substrate. This can be realized by physical vapor deposition of Ge atoms on a crystal surface. The formation of germanene has been first reported on Au(111) [3,4] and Pt(111) [5]. The growth of germanene has also been attested on Al(111) [6–9], Ag(111) [10,11], h-AlN [12], and on lamellar substrates such as $MoS_2$ [13] and highly oriented pyrolytic graphite (HOPG) [14]. Germanene films were also claimed to be obtained by segregation of germanium through a thick Ag/Ge(111) deposit caused by high temperature annealing [10,15,16], and on the surface of $Ge_2Pt$ droplets formed by high temperature annealing of a thin Pt/Ge(110) deposit [17].

However, some of these results have been recently questioned. On HOPG, it has been shown that the corrugation observed after Ge deposition was due to a charge density wave, and not to germanene [18]. On Au(111), a recent study concluded to the absence of germanene formation in a large temperature range (297 K- 500 K) [19]. For this metallic substrate, the observed structures would in fact correspond to a surface Au-Ge alloy [19]. The hypothetical formation of a surface alloy has also been recently discussed for Ge/Al(111) [20,21]. Similar questions may arise for germanene on Pt(111) due to the existence of a stable $Ge_2Pt$ bulk phase.

The case of germanene growth on Ag(111) is also controversial: some authors describe the result of depositing Ge on Ag(111) as an alloy, while others claim there is formation of a germanene single layer. In the following, the germanium coverage $\theta_{Ge}$ is given with respect to the atomic density of a Ag(111) plane; the coverage needed for a germanene monolayer (ML) is then $\theta_{Ge} = 1.06$ ML, assuming a germanene lattice constant of 0.397 nm [2]. The first study has been performed by depositing $\theta_{Ge}=1/3$ ML at room temperature [22]. After deposition, low energy electron diffraction (LEED) indicated the presence of a $p(\sqrt{3} \times \sqrt{3})R30°$ reconstruction related to the Ag(111) surface (hereafter named as $(\sqrt{3} \times \sqrt{3})$), but no contrast associated with this structure could be evidenced in the scanning tunneling microscopy (STM) images, which showed a $(1\times1)$ structure, similar to the one of bare Ag(111) [22]. X-ray photoelectron spectroscopy (XPS) revealed the metallic character of the Ge atoms in the $(\sqrt{3} \times \sqrt{3})$ superstructure, while from DFT computation, the authors concluded that the formation of a surface $Ag_2Ge$ alloy was favored with respect to adsorbed Ge adatoms. The model of $Ag_2Ge$ surface alloy was confirmed by angle-resolved photoelectron spectroscopy (ARPES), even if all observed features could not be reproduced by the simulation, in particular an unexpected surface band split at the -M points along the -Γ-K-M line of the $(\sqrt{3} \times \sqrt{3})$ surface Brillouin zone (SBZ) [23]. Annealing the $(\sqrt{3} \times \sqrt{3})$ structure at 473K resulted in a $(\sqrt{3} \times 6\sqrt{3})$ LEED pattern, corresponding to satellite spots around the initial spots of the $(\sqrt{3} \times \sqrt{3})$ reconstruction, and to a striped pattern in the STM images due to a long range modulation [24]. Contrary to the previous observation [22], the local $(\sqrt{3} \times \sqrt{3})$ structure was clearly seen in the STM images, with protrusions that could correspond to Ge atoms in the model of $Ag_2Ge$ surface alloy, with a small distortion. It was thus concluded that the complex surface band structure, with several split bands, most likely originates from the structural distortions of the alloy layer [24]. More recently, other STM investigations during growth at 300 K showed clear evidence of Ge substitution in the very initial stage of deposition, followed by the formation of a triangular pattern where amorphous-like and $(\sqrt{3} \times \sqrt{3})$ phases alternate [25]. For deposition at 600 K, the formation of the $(\sqrt{3} \times 6\sqrt{3})$ striped pattern was confirmed and also interpreted as the substitutional Ag$_2$Ge surface alloy [25].

In contradiction with these reports, some studies have concluded to the formation of germanene after Ge deposition on Ag(111). A STM and LEED study questioned the alloy model for the $(\sqrt{3} \times \sqrt{3})$ reconstruction [26]. Other authors observed the $(\sqrt{3} \times 6\sqrt{3})$ striped phase for deposition at 423 K, but for a coverage of 0.74 ML, which was interpreted as germanene with a large tensile strain of 12% to 23 % as a function of the direction [11]. In the same study, for a larger deposit of about 1.08 ML, this striped phase was replaced by a "quasi free-standing germanene phase", also identified as an honeycomb germanene layer, but with a slight compressive strain and with slight disorder [11]. This transition was confirmed in another study [27] where the structure that replaced the $(\sqrt{3} \times 6\sqrt{3})$ striped phase was called "disordered honeycomb". However, it was interpreted as the replacement of all Ag atoms of the Ag$_2$Ge alloy by Ge atoms.

Other studies proposed alternative structures for deposits above 1/3 ML. At 300K, the early works indicated that the $(\sqrt{3} \times \sqrt{3})$ superstructure was replaced by a centered-rectangular structure of size $c(\sqrt{3} \times 7)$, that has been associated with a coverage of 16 Ge atoms (for 14 Ag surface sites), forming 4 tetramers [28]. At 415K, another group [26] observed different phases for increasing amounts of Ge: $(9\sqrt{3} \times 9\sqrt{3})R30$, $c(\sqrt{3} \times 7)$ and $(12 \times 12)$, forming moiré structures, the first one being identified as germanene, with a tensile strain of about 13% with respect to free-standing germanene [2]. Finally, it was proposed that the further deposition of Ge leads to the growth of few-layers germanium films, terminated with a germanene layer displaying a $(\sqrt{3} \times \sqrt{3})R30°$ reconstruction related to Ge(111) [27].

To summarize, similar results obtained by the same techniques have been differently interpreted. The $(\sqrt{3} \times 6\sqrt{3})$ striped phase has been obtained either for 1/3 ML or for 0.74 ML, and interpreted either as a substitutional Ag$_2$Ge alloy, or as a very highly strained honeycomb germanene layer. For increasing Ge amounts, different structures, either well or badly ordered, have been also interpreted as germanene layers, but with different amounts of Ge related to compressive or tensile strains, or without strain. The contradictions between these STM results show that they suffer from a lack of real-time observation of the surface, performed at the growth temperature and during the growth. This investigation should indeed permit to distinguish the processes in play and to determine the nature of the different Ge/Ag(111) structures, in relation with the amount of Ge. Such studies have previously been shown to be powerful for following the growth of Si layers on Ag(110) and on Ag(111) [29–31]. In the present study, we have followed in real-time by STM the surface evolution upon Ge deposition around 400 K on Ag(111). From the evolution of the Ag(111) step edges, we determined that the $(\sqrt{3} \times \sqrt{3})$ reconstruction is a Ag$_2$Ge surface alloy, and that the disordered hexagonal structure also corresponds to a surface alloy, with a larger proportion of Ge atoms.

2. Experimental

Sample preparation and STM experiments have been performed in an ultra-high vacuum (UHV) system with a base pressure less than $1 \times 10^{-10}$ mbar equipped with an Omicron variable temperature STM. The Ag(111) single crystal was prepared by several cycles of Ar ion sputtering ($P = 7 \times 10^{-5}$ mbar, 600 eV) and annealing ($T = 870$ K). Germanium was evaporated from a graphite crucible using an Omicron Nanotechnology e-beam evaporator installed in front of the STM with a very low flux $F \approx 0.1$ ML/h. All coverages have been calibrated using the completion of the striped phase as a reference, set to a coverage of 1/3 ML.

STM imaging was performed in real-time during Ge evaporation, with a substrate held at a fixed temperature in the [380 K - 430 K] range. The pressure was always below $1\times10^{-10}$ mbar during the experiments. Depending on the tip shape, small or large shadowing effects can occur during deposition. Indeed, as the STM tip remains very close to the surface (a few Å), the area located behind the tip with respect to the direction of the incoming flux is not covered by Ge atoms. However, as the tip moves over the surface while scanning, the shadow also moves, which gives rise to local inhomogeneities on the surface. In order to check that the growth is not influenced by the tip, we have performed experiments for which the tip was removed during evaporation, in order to avoid shadowing effects. For all cases, comparison between images of the same area performed at different times have been made by carefully correcting the STM images from the drift using a home-made procedure [30].

3. Results

3.1 Evolution of the Ge/Ag(111) structures

Fig. 1 shows the evolution of the same area during continuous Ge evaporation on a Ag(111) surface held at 380K. It corresponds to a detailed view (480 x 480 nm$^2$) of the larger area (680 x 680 nm$^2$) continuously scanned during Ge evaporation (See Movie in the Supplemental Material [32] for the complete film). Due to shadowing effects, the Ge flux is higher on the top left side of the image than on the bottom right side. Before evaporation, the surface displays flat terraces separated by single atomic steps. The colors correspond to terraces at different levels, the lower ones being at the up-right part of the images. As imaging is performed during evaporation, Ge atoms also adsorb on the STM tip, which can lead to noisy parts on the STM images. The dark triangle at the center of the image is a defect that serves as a marker to follow the surface evolution. Note that step edges are also pinned at atomic defects of the surface. Whereas before Ge deposition, atomic steps display only small fluctuations, during deposition, the step edges are mobile and move towards the right part of the image, i.e. toward the descending direction, forming outgrowths on the inferior terraces, as shown by the white arrows in Fig.1.b. Such behavior has already been observed during Si deposition on Ag(110) [29] and on Ag(111) [30]. For Ag(110), it has been explained by the formation of a Ag missing-row reconstruction during which released silver atoms form outgrowths from the step edges. For Ag(111), it has been explained by the exchange between Si and Ag atoms in the surface plane, leading to the release of Ag adatoms that further condense at the step edges or grow new terraces. For both cases, the outgrowths and new terraces are initially free of silicon, and then covered with Si, as Si deposition continues.

In the present case, the situation is different. In Fig. 1b-1d, where outgrowths have developed at step edges, especially in the top of the images, the presence of Ge on the surface is also clearly visible. This is in particular the case in Fig. 1c, where dark twig-like dendrites appear (see Fig. S2a [32] for a detailed view), or in the bottom right part of Fig. 1d, where triangles have formed (see Fig. S6a [32] for a detailed view). As can be seen in Fig. S1 and S3, the Ge coverage is the same in the regions of the outgrowths and in the initial terraces, which indicates that the growing outgrowths are instantaneously (at the timescale of the observation) covered with Ge. This evolution has been observed for all substrate temperatures studied (see Fig. S4 for growth at 416 K).

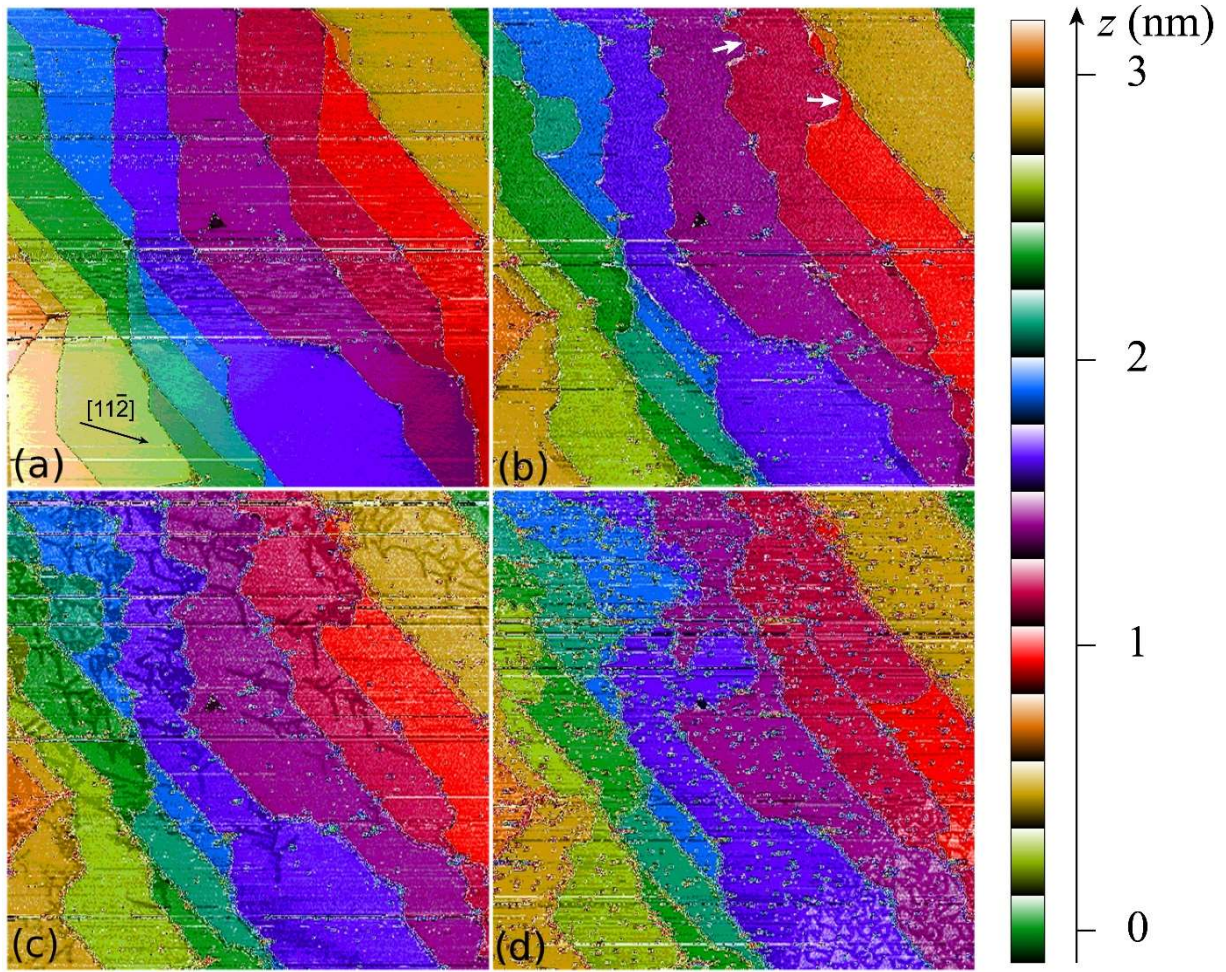

Fig. 1. Evolution of the Ag(111) surface during Ge deposition at 380 K, for (a) 0 min, (b) 100 min ($\theta_{Ge}$ = 0.08 ML), (c) 190 min ($\theta_{Ge}$ = 0.16 ML), (d) 360 min ($\theta_{Ge}$ = 0.30 ML). In (b), the white arrows indicate the outgrowths that have grown during Ge deposition. Size of the images: 580 × 580 nm$^2$. Tunneling conditions : $V_S$ = 1.4 V - $I$ = 30pA. See Movie in the Supplemental Material [32] for the complete film.

Fig. 2 displays detailed views of the surface evolution for different evaporation times. At the very beginning of the evaporation, some dark spots are visible on the STM images (Fig. 2a). They have been shown to correspond to Ge atoms in substitutional position [25] that occupy randomly the surface. Similar observations have been performed in the case of Si/Ag(111), after deposition of a very low Si coverage at room temperature [33].

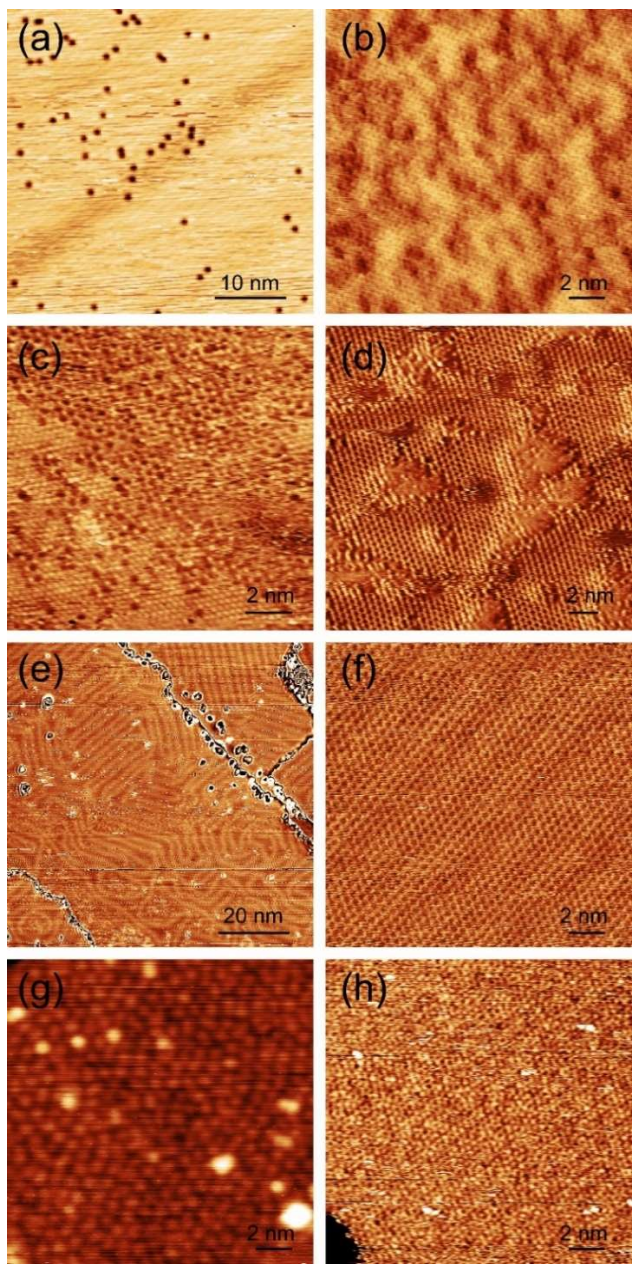

Fig 2. Detailed view of the Ge/Ag(111) surface for increasing coverage. a) diluted phase for $\theta_{Ge} = 0.0025$ ML. b) diluted phase at saturation, i.e. just before the formation of twig-like dendrites. c) boundary between the diluted phase and a dendrite. d) triangle phase. e-f) striped phase. g-h) disordered hexagonal phase. Deposition temperatures are 380 K (a, e), 400 K (d), 414 K (f-h), 420 K (b-c). Images (a) and (e) are acquired during deposition, the other images are acquired at 300 K after deposition. In (e), the image is displayed modulo the step height. Tunneling conditions: (a, g) $V_S = 1.5$ V-$I = 30$ pA; (b) $V_S = 0.4$ V-$I = 20$ pA; (c) $V_S = -0.3$ V $I = 0.5$ nA; (d) $V_S = 0.4$ V-$I = 0.2$ nA; (e) $V_S = 1.0$ V-$I = 50$ pA; (f) $V_S = 0.3$ V-$I = 0.2$ nA; (h) $V_S = 0.4$ V-$I = 0.1$ nA.

As deposition time increases, this dilute Ge phase becomes denser, Ge atoms form denser areas separated by nearly pure Ag regions, with a characteristic length of 2-3 nm (Fig. 2b).

A first order transition occurs for a coverage of ~0.11 ML, where twig-like dendrites form on the surface, as shown in Fig. 1c. A detailed view of the surface shows that they are roughly oriented along the <112> directions of Ag(111) (see Fig. S2a [32]). In the STM images, they appear as darker areas (i.e., at a lower apparent height). They display a $(\sqrt{3} \times \sqrt{3})$ structure, while the lighter areas in their vicinity have a $(1 \times 1)$ structure. This is clearly visible in Fig. 2c where a detailed view of the frontier between a $(1 \times 1)$ region at the bottom left part of the image and a $(\sqrt{3} \times \sqrt{3})$ region (upper right part) is shown. The height profiles along the main axes of the different phases (Fig. S5b) indicate a lattice constant of 0.29 nm and 0.52 nm for the $(1 \times 1)$ and $(\sqrt{3} \times \sqrt{3})$ regions, in reasonable agreement with the expected values for the $(1 \times 1)$ and $(\sqrt{3} \times \sqrt{3})$ lattice constants (0.289 nm and 0.500 nm respectively). The comparison between successive images (see Fig. S2a and S2b [32]) shows that the dendrites form by the assembly of three-pointed hollow stars.

As coverage increases, the network of twig-like dendrites becomes denser and transforms into a triangular network where the $(1 \times 1)$ domains of the diluted Ge phase have a triangular shape and are surrounded by the $(\sqrt{3} \times \sqrt{3})$ regions. This can be seen in the bottom right corner of Fig. 1d and in the detailed view shown in Fig. 2d. The height profiles along the main axes of the $(\sqrt{3} \times \sqrt{3})$ phase (Fig. S2d [32]) indicates a lattice constant of 0.50 nm. Note that these triangles adopt a more regular shape and form a more ordered network as the growth temperature increases (see Fig. S6 [32]).

For $\theta_{Ge} = 1/3$ ML, another sudden transition occurs where triangle domains are replaced with the striped phase (Fig. 2e, 2f and S7a [32]). The detailed view of Fig. 2e, where the coverage is slightly higher at the top of the image than at the bottom, shows that the striped phase initially follows the orientation of the initial twig-like dendrites, but very rapidly transforms into larger domains with a single orientation (the stripes are oriented along <112>). This demonstrates that this reorganization into large areas does not imply a large amount of material transport.

Finally, local patches of a disordered-hexagonal (DH) reconstruction form. They start to grow at step edges, as already reported [11], but they also form inside large terraces. Depending on the STM tip termination, the DH phase appear as a disordered hexagonal array of protrusions, with a lattice constant of 0.97±0.02 nm (Fig. 2g and S7d-f [32]), or as a more organized hexagonal array, with a smaller lattice constant of 0.43±0.02 nm (Fig. 2h and S7g-i). This is significantly higher than the value found previously (0.391 nm) [11], and also higher than the lattice constant of free-standing germanene (0.397 nm) [2]. From the analysis of the evolution of the fraction of each phase as a function of coverage, we have estimated that this DH reconstruction is completed for 0.6±0.1 ML. This is far from the coverage $\theta_{Ge} = 1.06$ ML needed for a germanene monolayer. This disordered hexagonal layer must thus contain a significant amount of Ag atoms.

For higher coverage, a second layer starts to form above the disordered hexagonal layer, as already reported. [27]

In Fig. 3 is plotted the surface fraction of each structure as a function of Ge coverage. The values have been obtained from the analysis of the successive images obtained during Ge deposition at 380 K (and corresponding to the movie [32]). The succession of the different phases is clearly visible. As there is a gradient of coverage in each of the successive recorded STM images, the transitions between the different structures are smoothed. The extrapolation of the linear growth of the DH phase is also in a good agreement with a completion of this phase for $\theta_{Ge} \approx 0.6$ ML.

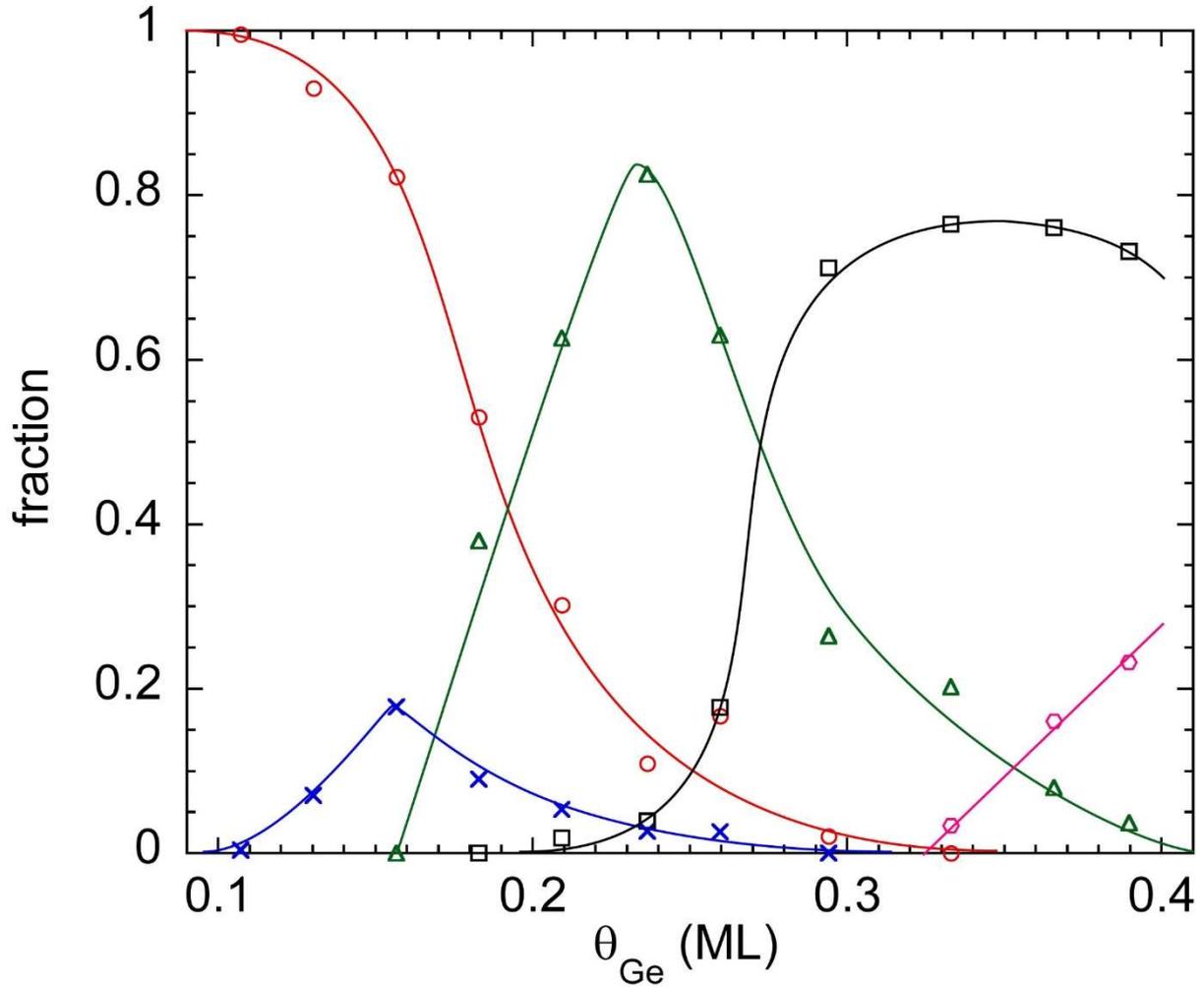

Fig. 3. Fraction of the different phases as a function of Ge coverage during growth at 380K. Red dots: diluted phase, blue crosses: twig-like dendrites, green triangles: triangles, black squares: striped phase SP, pink hexagons: disordered hexagonal (DH) phase. Lines are guides for the eyes. Note that although consisting in the mixing of two phases, the triangle phase is counted as a whole.

3.2 Ag content in the structures

In order to determine the fraction of Ag atoms in the observed structures, we have precisely measured the advance of Ag step edges, which corresponds to the quantity of Ag atoms that are expelled by Ge atoms during the deposition. The Ag concentration in the observed structures is thus the complement of the concentration of expelled Ag atoms ([32]). The dependence of Ag concentration with Ge coverage, measured on STM images obtained during Ge deposition at 380 K (see Fig. 1 and movie [32]) is drawn in Fig. 4.

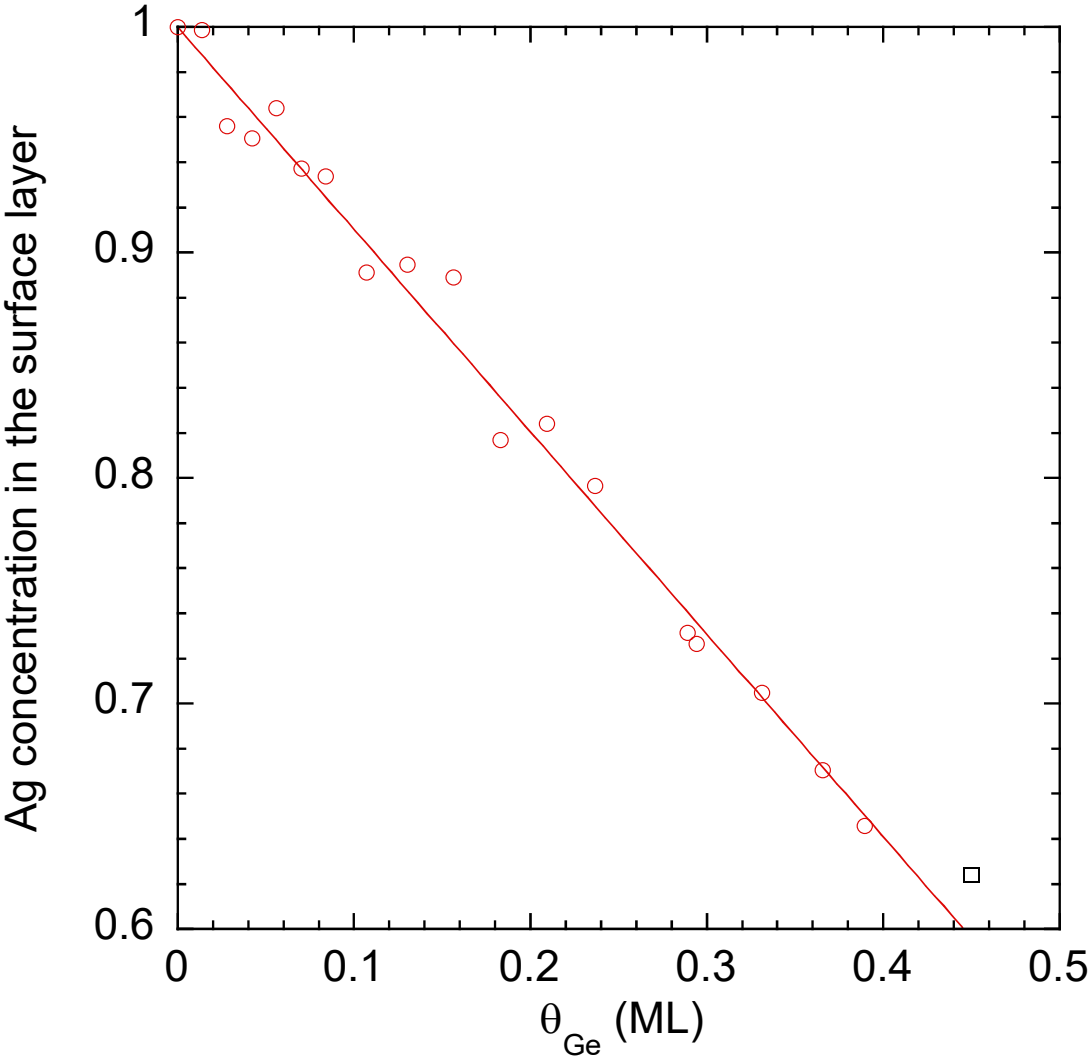

Fig. 4. Evolution of the Ag concentration in the surface structures, as a function of Ge coverage, during deposition at 380K. Red circles correspond to the experiment reported in Fig.1. The line is a linear fit with slope -0.9. The black square corresponds to another deposit at the same temperature, on an area not affected by shadowing effect.

The figure shows that the Ag concentration and Ge coverage are related through a linear relationship, where 1 Ge atom replaces around 0.9±0.1 Ag atom. Note that this latter quantity could be slightly underestimated due to the formation of some Ag clusters on the surface, which can be seen by STM during deposition. As it is not possible to determine whether they contain Ag or Ge, we have excluded them from the measurements. Taking into account the uncertainty related to the measurement, this is in correct agreement with a $Ag_2Ge$ surface alloy for the $(\sqrt{3} \times \sqrt{3})$ structure obtained for 1/3 ML. Concerning the DH structure, the extrapolation of the data presented in Fig. 4 indicates that the completion should be obtained for a Ge coverage nearly equal to twice the coverage for the $(\sqrt{3} \times \sqrt{3})$ structure, i.e., 0.6±0.1 ML, largely smaller than the expected coverage of 1.06 ML for germanene. This DH structure is thus an Ag-Ge alloy and as the decrease of Ag coverage in the structures is linear (Fig. 4), its extrapolation to higher Ge amounts indicates that the DH structure must contain a significant amount of Ag, i.e. 0.46±0.09 ML. This is compatible with a $Ag_3Ge_4$ surface alloy. Due to the large error bars associated with the measurements, the DH structure is also compatible with AgGe or $AgGe_2$ alloys. However, from this analysis as a function of the actual amount of Ge in these two structures, we can conclude with good confidence that, in both cases, they are not germanene layers, but rather Ag-Ge alloys with different concentrations in Ge.

3.3 Diffusion of Ge atoms

In the temperature range of our experiments, the inserted Ge atoms that are observed for very low coverage (i.e $\theta_{Ge} < 0.001$ ML) are quite mobile. At 380 K, we observe that more than half of the Ge atoms have moved, from one atomic site to another, during the time interval of 13 min between two consecutive images (see Fig. S8a-c). We have analyzed the displacements of Ge atoms between consecutive images obtained at this temperature. In the analysis, we have only taken into account Ge atoms far from the step edges, and having not too close neighbors.

We have also excluded Ge displacements higher than $10d_0$, where $d_0 = 0.289$ nm is the Ag-Ag interatomic distance, since it becomes comparable to the average interatomic distance between Ge atoms. Fig. 5a shows the histogram of Ge displacements, normalized to $d_0$, measured on the trajectories of 866 Ge atoms. About 32% of the atoms have not moved from one image to the next one, while 39% have moved by one atomic position. Some atoms have travelled a larger distance, and the mean displacement is $d = 1.26 d_0$.

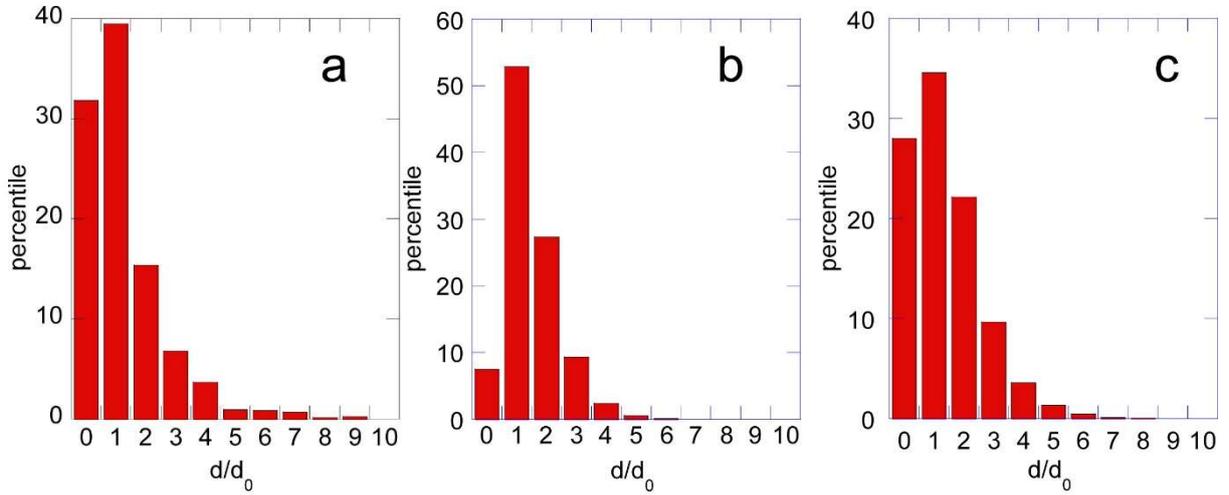

Fig. 5. Histograms of Ge atomic displacements, normalized to the to the Ag-Ag interatomic distance. a) measured between successive images acquired at 13 min interval, and at T=380 K. b) best fit within the framework of single diffusion mechanism. c) best fit within the framework of diffusion through deinsertion, jumps and reinsertion.

Several atomistic mechanisms may be involved in surface diffusion of atomic species [34]. As Ge atoms are inserted in the surface plane, either they diffuse in the surface plane, or above the surface. The first case corresponds to vacancy-assisted diffusion. It has been proposed for example for In atoms inserted in Cu(100) [35]. In that case, Ge atoms jump into a nearest neighbor site by exchange with the vacancy. Two possibilities can be considered. If the vacancy is mobile and does not remain in the vicinity of the Ge atom, the next jump will occur with another vacancy. Diffusion corresponds thus to a series of uncorrelated jumps over a distance $d_0$. We have computed the histogram of displacements associated to such mechanism, using a basic Monte-Carlo simulation program of atomic diffusion (see Supplemental Material for a detailed description of the simulations). With this model, it is not possible to correctly reproduce the histogram of atomic displacements. The distribution, shown in Fig. 5b, is more peaked around the average value, and the fraction of atoms that have covered a distance larger than $5d_0$ is negligible.

On the contrary, if the vacancy remains in the neighborhood of the inserted atom, long correlated jumps may occur. This is generally the case for large atoms inserted in a surface, as In in Cu(100), where the atom-vacancy interaction is attractive since the presence of the vacancy reduces the compressive stress associated with the insertion of the foreign atom. In this case, periods of long jumps, when the vacancy remains close to the inserted atom, are separated by periods where the inserted atoms keep still. In the present case, the atomic radius of Ge is smaller than the one of Ag, so we expect a repulsive interaction between an inserted Ge atom and a vacancy, which makes the vacancy-assisted diffusion mechanism improbable.

The second mechanism of diffusion involves a three-step process: exchange between an Ag adatom and a Ge inserted atom, diffusion of the Ge atom on the surface and reinsertion through exchange with a surface Ag atom. It has been for example shown to describe the diffusion mechanism of Pb/Cu(110) [36]. For this mechanism, the activation energy for deinsertion determines the frequency of diffusion events, $\nu_{dis} = \nu_{dis}^0 \exp\left(-\frac{E_{dis}}{kT}\right)$, whereas the difference in activation energy between atomic jumps on the surface and reinsertion determines the average length of a diffusion event. We note $p_{ins} = \dfrac{\nu_{ins}^0 \exp\left(\dfrac{-E_{ins}}{kT}\right)}{\nu_{ins}^0 \exp\left(\dfrac{-E_{ins}}{kT}\right) + \nu_{jump}^0 \exp\left(\dfrac{-E_{jump}}{kT}\right)}$ the relative probability of reinsertion of a Ge atom. We have simulated the diffusion of Ge atoms with such a mechanism. A good agreement is obtained with a characteristic frequency for deinsertion $\nu_{dis} = 2 \; 10^{-3}$ s$^{-1}$, and a relative probability of reinsertion of 0.4, indicating that the difference in energy barrier between insertion and jump is small, of the order of 0.03 eV, for a ratio between $\nu_{ins}^0$ and $\nu_{jump}^0$ not larger than 10. The corresponding histogram is shown in Fig. 5c and is in excellent agreement with the experimental histogram of Fig. 5a. This suggests that the exchange mechanism is easier for Ge/Ag(111) than for the similar system, Si on Ag(111), for which the difference in activation energy between atomic jump and exchange was shown to be 0.17eV [37].

For an attempt frequency $\nu_{dis}^0 = 10^{12}$ s$^{-1}$, we obtain an activation energy for deinsertion $E_{dis} = 1.1$ eV. Note that this value includes the Ag adatom creation energy, which value has been calculated from DFT: $E_{ad} = 0.67$ eV [29]. Thus, the activation barrier for exchange strictly speaking is of the order of 0.4 eV only. At 300 K, with $E_{dis} = 1.1$ eV, we obtain that $\nu_{dis} = 3.10^{-7}$ s$^{-1}$ and no atomic motion can be observed at the timescale of the experiment. On the contrary, at 415 K, we derive $\nu_{dis} = 0.03$ s$^{-1}$. We have followed the diffusion at this temperature on images scanned with the up and down mode, with a scan speed of 0.75s/line. On the very top of the images, where similar lines on two consecutive images are scanned with a small time interval, some atoms are observed at the same position (Fig S8d-e [32]). As the time interval between similar lines increases, the proportion of atoms that have moved increases. After a time interval of 300 s, all atoms are observed at different position. This is in good agreement with the diffusion simulations performed at this temperature, that indicate that for these parameters, only 1% of the atoms should be found at the same position.

Conclusion

In summary, we have followed by real-time STM the evolution of the successive structures formed on Ag(111) upon Ge deposition in the [380 K - 430 K] temperature range. Our results show that Ge atoms easily exchange with Ag atoms to form various Ag-Ge surface alloys. For 1/3 ML, the substitutional $Ag_2Ge$ alloy forming a striped phase is observed. For a coverage of 0.6 ML, the disordered hexagonal structure is obtained, which is also interpreted as an alloy with a larger proportion of Ge atoms. No structure was found compatible with a germanene layer. The quantitative analysis of diffusion of isolated Ge atoms shows that atoms move in a three-step process: exchange between an Ag adatom and a Ge inserted atom, diffusion on the surface of the Ge atom and reinsertion through exchange with a surface Ag atom. The activation energy for Ge/Ag exchange is small, which leads to a high mobility of Ge atoms in the [380 K - 430 K] temperature range studied.

Acknowledgments

This study is financially supported the French National Research Agency (Germanene project ANR-17-CE09-0021-03). K. Z. is supported by the Chinese Scholarship Council (CSC contract 201808070070).

Growth of germanium-silver surface alloys followed by in-situ scanning tunneling microscopy: absence of germanene formation.

K. Zhang,[1] R. Bernard,[1] Y. Borensztein,[1] H. Cruguel,[1] G. Prévot[1]

[1]Sorbonne Université, CNRS, Institut des NanoSciences de Paris, INSP, F-75005, Paris, France

Supplemental Material

**Determination of the Ag content in the structures.**

In this paragraph, we show that the fraction of Ag atoms in the observed structures is the complement of the advance of Ag step edges. For the demonstration, we can start with a planar surface. Before deposition, the whole surface is at *z=0*. The Ag surface coverage is the Ag content in the surface plane and is equal to 1. After deposition of a given Ge amount, expelled Ag atoms form outgrowths that occupy a fraction *x* of the surface. The surface plane of these outgrowths, with a Ag coverage $\theta_{Ag} < 1$, is now at *z=d,* where *d* is the Ag(111) interlayer spacing. Below these outgrowths, the plane at *z=0* is a pure Ag plane, whereas in the rest of the surface, the plane at *z=0* has a coverage $\theta_{Ag}$ as it is covered with the same Ag-Ge structure as the outgrowths.

The conservation of Ag atoms leads to $1 = (1 - x)\theta_{Ag} + x(1 + \theta_{Ag})$. Thus, $\theta_{Ag} = 1 - x$.

**Monte-Carlo simulations**

Single atom trajectories were computed using a Monte-Carlo program based on the Metropolis rejection algorithm. Two models were tested.

In the first one, Ge atoms diffuse in the surface plane into vacant sites. Ag vacancies are treated in a mean-field approach and the probability that one of the six nearest neighbour sites is vacant is proportional to $e^{-\frac{E_{vac}}{kT}}$ where $E_{vac}$ is the vacancy formation energy. The diffusion probability per unit time is $v_{vac}^0 e^{-\frac{E_{bvac}+E_{vac}}{kT}}$ where $v_{vac}^0$ is a characteristic frequency and $E_{bvac}$ is the energy barrier for Ge-vacancy exchange.

In the second one, the considered diffusion events were atom disinsertion, jumps and reinsertion. Disinsertion events occur through exchange between Ag adatoms and Ge inserted atoms, with an energy barrier $E_{bdis}$. Ag adatoms are treated in a mean-field approach and the probability that a Ag adatom is next to a Ge inserted atom is proportional to $e^{-\frac{E_{ad}}{kT}}$ where $E_{ad}$ is the Ag adatom formation energy. The probability of disinsertion per unit time is $v_{dis} = v_{dis}^0 \exp\left(-\frac{E_{dis}}{kT}\right)$ where $E_{dis} = E_{bdis} + E_{ad}$. Disinserted atoms either diffuse by single atomic jumps with probability $p_{jump}$, or reinsert in the surface with probability $p_{ins}$. As no other atomic diffusion processes are considered, $p_{ins} + p_{jump} = 1$ and $p_{ins} = \frac{v_{ins}}{v_{ins}+v_{jump}}$ with $v_{ins} = v_{ins}^0 \exp\left(\frac{-E_{ins}}{kT}\right)$ and $v_{jump} = v_{jump}^0 \exp\left(\frac{-E_{jump}}{kT}\right)$. Moreover, we have considered that $v_{ins} \gg v_{dis}$ and $v_{jump} \gg v_{dis}$ since the disinsertion frequency $v_{dis}$ includes the extra Ag adatom concentration. Histograms of Ge atomic displacements were computed on 10000 individual trajectories.

**Movie 1.**

Evolution of the Ag(111) surface during Ge deposition at 380K. Movie length: 7h25min. Size of the images: 500 × 500nm$^2$.

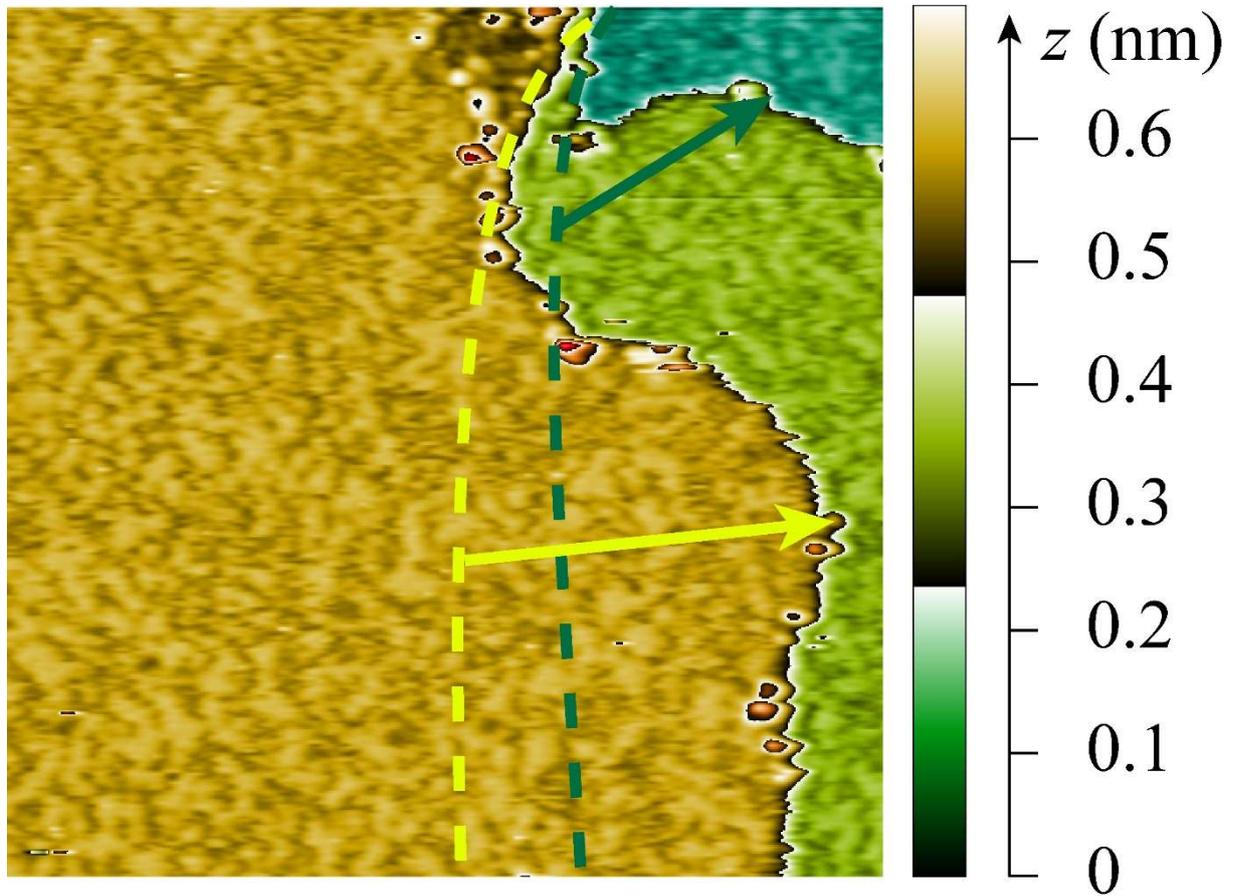

Fig. S1: Evolution of the Ag(111) surface during Ge deposition at 380K. The image shows the formation of outgrowths. The Ge coverage ($\theta_{Ge} = 0.1$ ML) is the same in the outgrowths and in the initial terraces. The position of step edges before evaporation is indicated by the dotted lines, and the motion of the step edges are given by the arrows. Size of the image: 70 × 70 nm². Tunneling conditions : $V_S$ = 1.4 V - $I$ = 30pA.

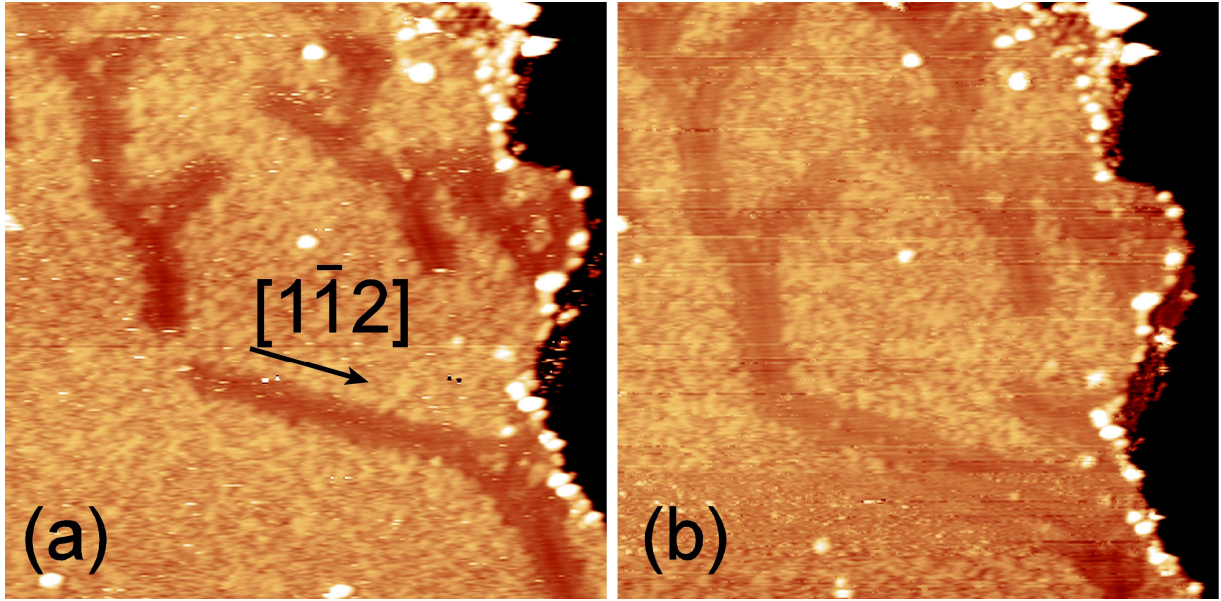

Fig. S2: Evolution of the Ag(111) surface during Ge deposition at 380K. Consecutive images (65 × 65 nm$^2$) acquired at 5 min time interval. The images show the formation of dark twig-like dendrites, oriented along the <112> directions. Tunneling conditions : $V_S$ = 1.4 V - $I$ = 30pA.

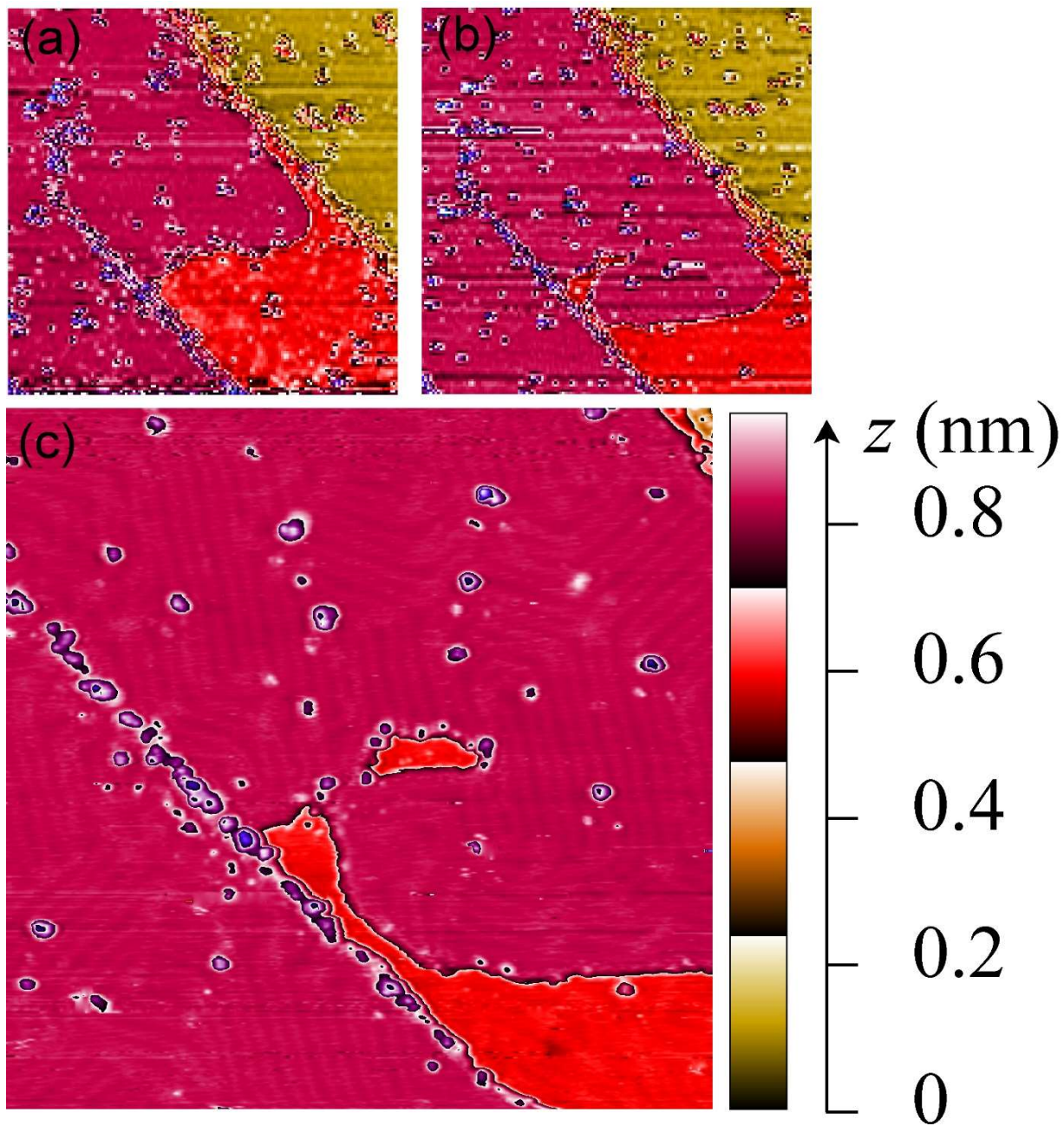

Fig. S3: Evolution of the Ag(111) surface during Ge deposition at 380K. (a) and (b): consecutive images (170 × 170 nm$^2$) acquired at 40 min time interval showing the motion of the outgrowth edge. The detail (c) show that the outgrowth is uniformly covered with the striped phase, as in the initial terrace. Tunneling conditions : $V_S$ = 1.5 V- $I$ = 30pA.

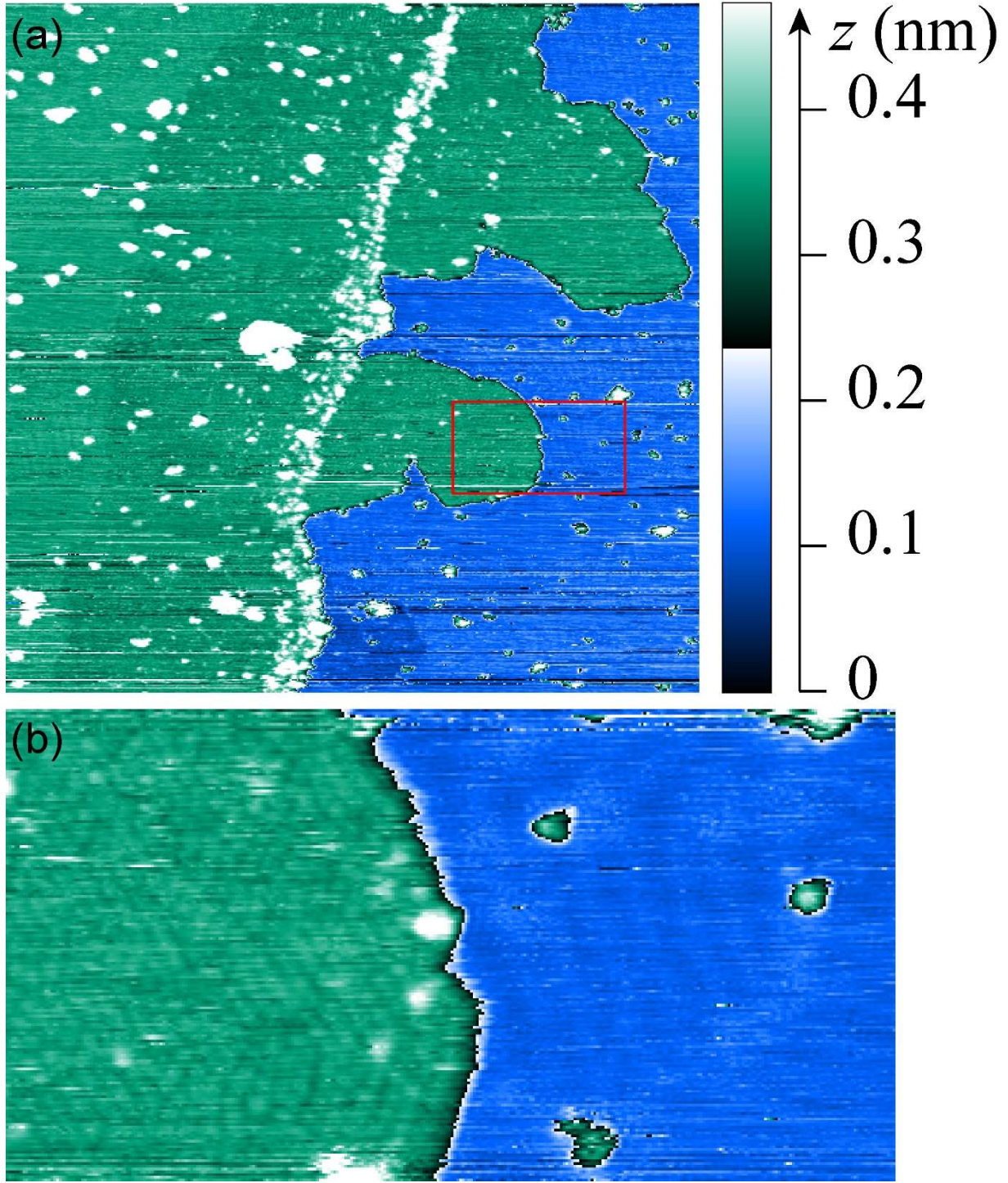

Fig. S4: Ag(111) surface during Ge deposition at 416K for an average coverage $\theta_{Ge} \approx 0.4$ML. (a) shows the formation of outgrowths at the right of the initial step edge position (marked by the presence of a line of clusters). The lower terrace (blue color) displays mainly the striped phase, while the upper terrace (green color) displays both the striped (at the left of the image) and disordered hexagonal phases (at the left of the initial step edge and in the outgrowths). Size of the image: 170x170 nm². The detailed view (b) corresponds to the red rectangle in (a). Tunneling conditions: $V_S = 1.6$ V - $I = 80$pA.

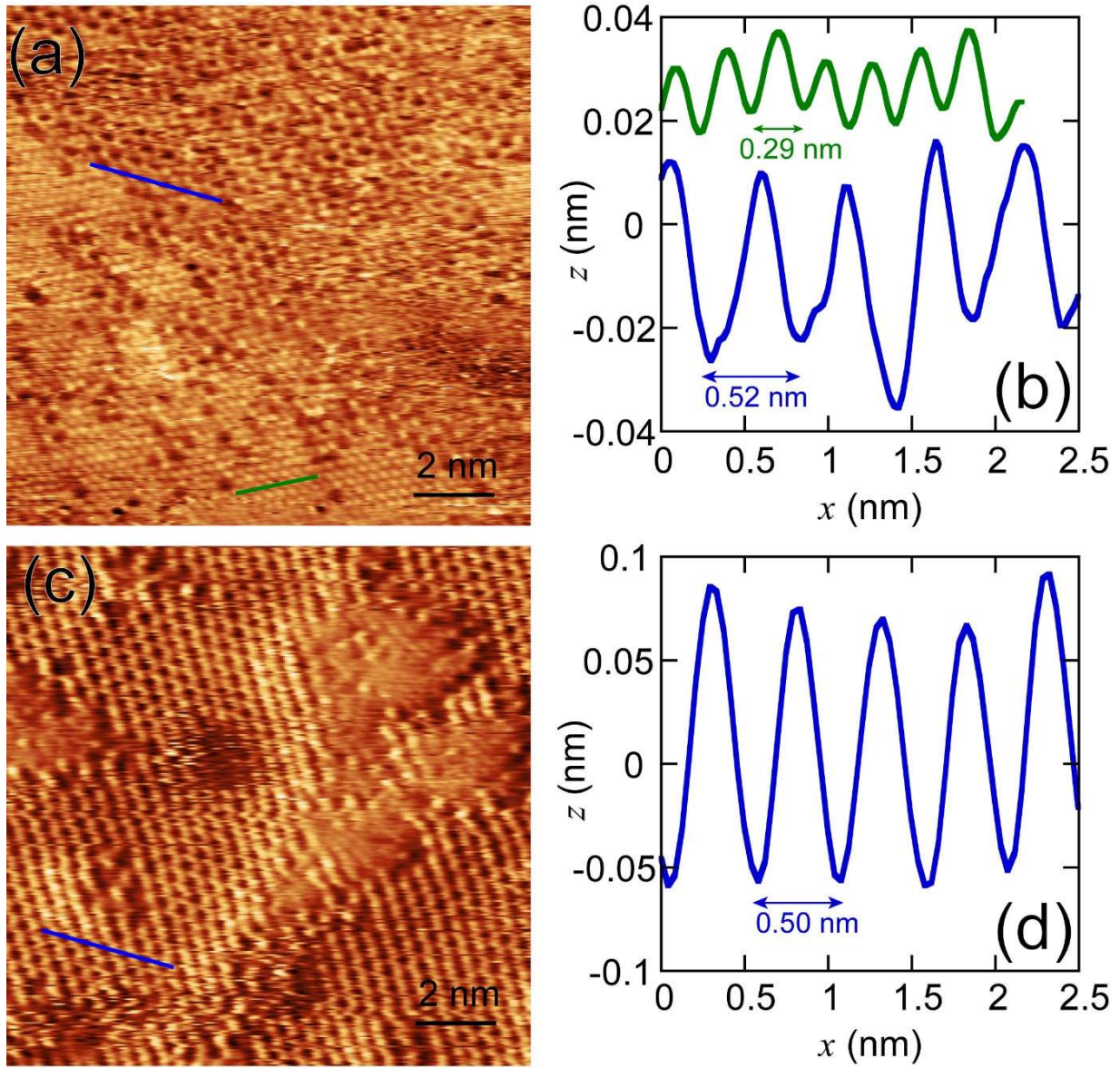

Fig. S5: a) Boundary between the diluted phase and a twig-like dendrite obtained after Ge deposition at 420 K. b) Height profiles along the lines drawn in a). c) Triangle phase obtained after Ge deposition at 400 K. d) Height profile along the line drawn in c). Tunneling conditions (a) -0.3 V $I$ = 0.5 nA; (c) $V_S$ = 0.4 V - $I$ = 0.2 nA;

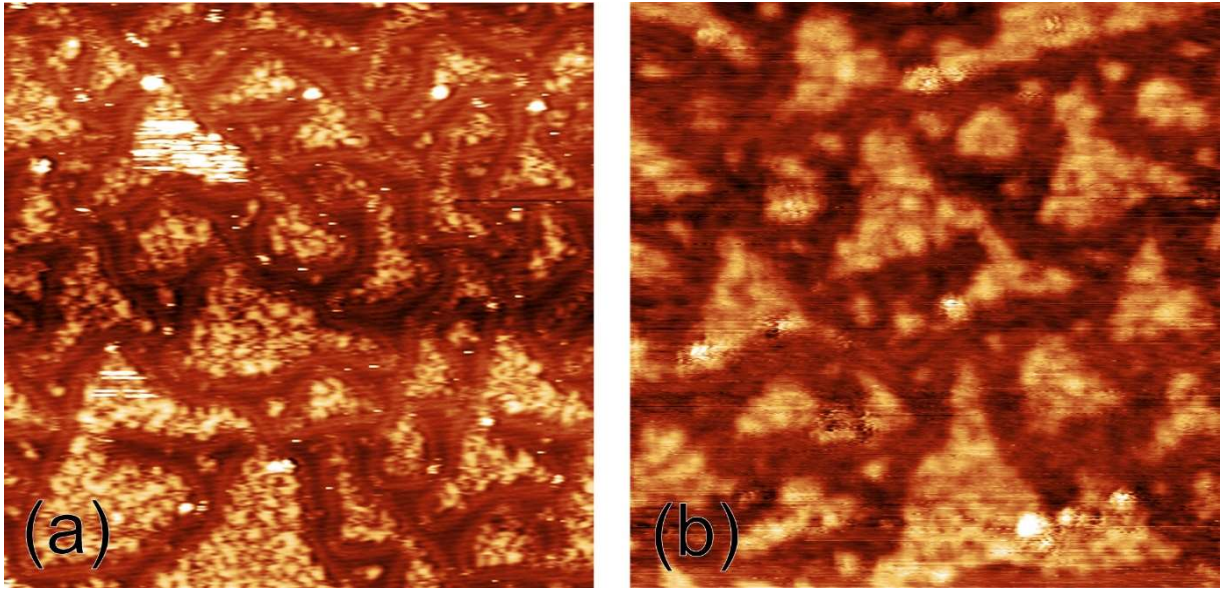

Fig. S6. Triangular network formed on Ag(111) surface after deposition of Ge at 380K (a) and 401 K (b). Size of the images (68 × 68 nm²). Tunneling conditions (a) 1.4 V $I$ = 30 pA; (b) $V_S$ = 0.4 V- $I$ = 0.1 nA;

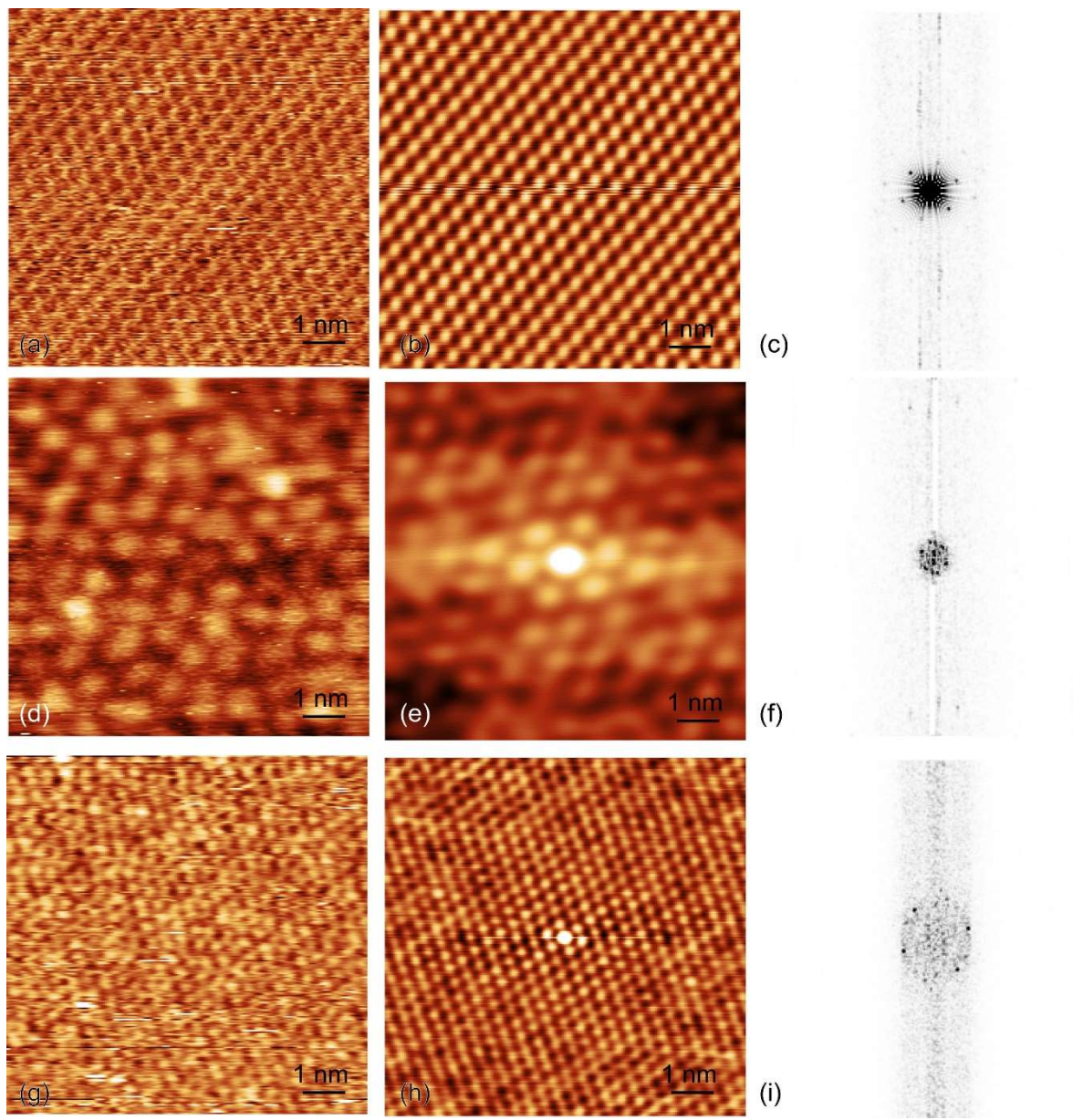

Fig S7. (a,d,g): STM images of the Ag(111) surface acquired at 300 K after Ge deposition at 414 K. (b,e,h): Corresponding self-correlation images. (c,f,i): Corresponding Fast Fourier Transform images. (a-c) show the striped phase, (d-f) and (g-i) show the disordered hexagonal phase. Images (d) and (g) were acquired with a different tip termination. Tunneling conditions (a) $V_S$ = 0.3 V - $I$ = 0.2 nA; (d) 1.5 V $I$ = 30 pA; (g) $V_S$ = 0.4 V - $I$ = 0.1 nA.

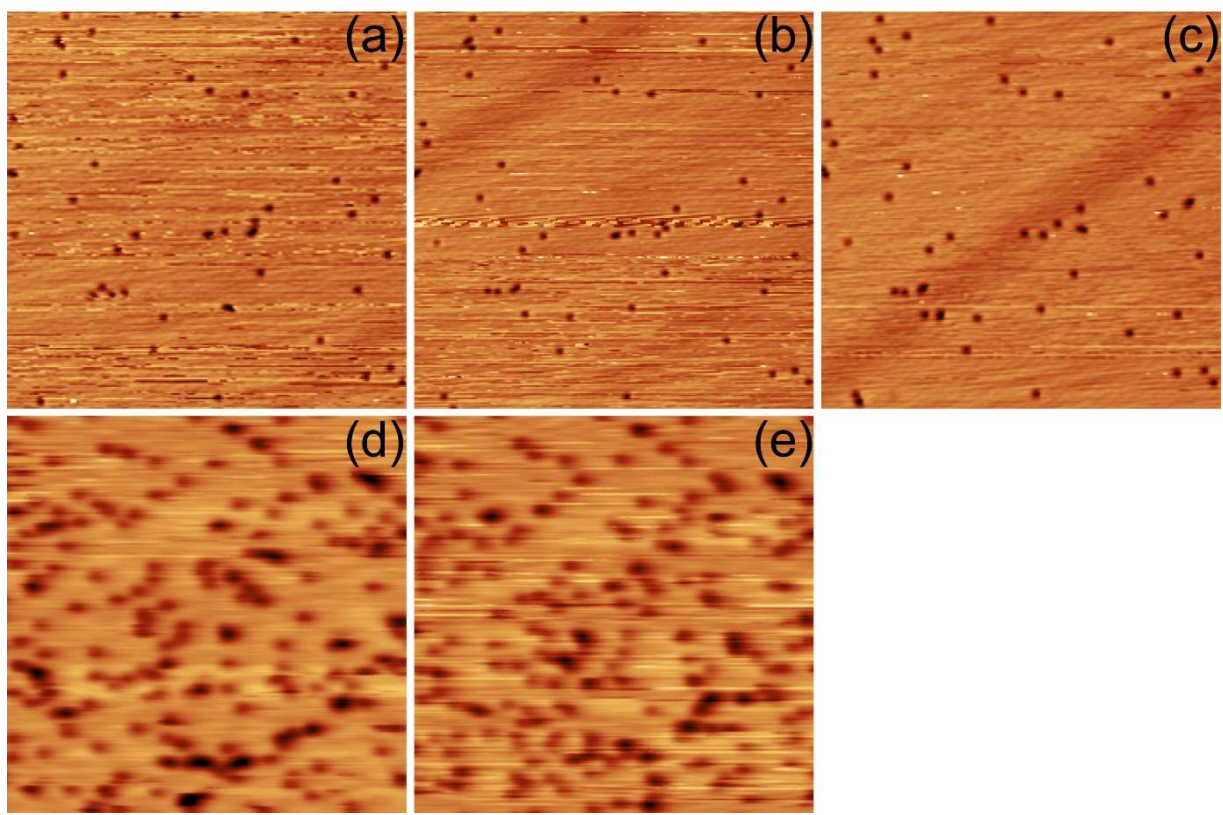

Fig. S8. (a-c). Diffusion of Ge atoms on Ag(111) at 380 K. Consecutive images recorded at a time interval of 13 min. (d-e). Diffusion of Ge atoms on Ag(111) at 415 K. Successive images acquired from the bottom to the top and from the top to the bottom at a scan speed of 0.75s/line. Size of the images: 43 x 43 nm$^2$. Tunneling conditions : $V_S$ = 1.5 V- $I$ = 30 pA.